\documentclass[onecolumn,journal]{IEEEtran}

\usepackage{amsmath,bbold, setspace, stfloats, dsfont,yfonts,amssymb,dsfont,epsfig,psfrag,amsthm,bm,multirow, graphicx,color,mathrsfs}

\usepackage{algpseudocode}% http://ctan.org/pkg/algorithmicx
\usepackage{algorithm,eqparbox,array}% http://ctan.org/pkg/algorithm
\usepackage{xspace}% http://ctan.org/pkg/xspace

\usepackage{caption2}

\newtheorem{proposition}{Proposition}

\newtheorem{lemma}{Lemma}

\begin{document}

\title{Scheduling in Time-correlated Wireless Networks with Imperfect CSI and Stringent Constraint}

\author{\emph{Wenzhuo Ouyang, Atilla Eryilmaz, and Ness B. Shroff}

\thanks{Wenzhuo Ouyang and Atilla Eryilmaz are with the Department of ECE, The Ohio State University (e-mails: ouyangw@ece.osu.edu, eryilmaz@ece.osu.edu).
Ness B. Shroff holds a joint appointment in both the Department of ECE and the Department of CSE at The Ohio State University (e-mail: shroff@ece.osu.edu).}
}
\maketitle

\begin{abstract}
In a wireless network, the efficiency of scheduling algorithms over time-varying channels depends heavily on the accuracy of the Channel State Information (CSI), which is usually quite ``costly'' in terms of consuming network resources. Scheduling in such systems is also subject to stringent constraints such as power and bandwidth, which limit the maximum number of simultaneous transmissions. In the meanwhile, communication channels in wireless systems typically fluctuate in a time-correlated manner. We hence design schedulers to exploit the temporal-correlation inherent in channels with memory and ARQ-styled feedback from the users for better channel state knowledge, under the assumption of Markovian channels and the stringent constraint on the maximum number of simultaneously active users. We model this problem under the framework of a Partially Observable Markov Decision Processes.

In recent work, a low-complexity optimal solution was developed for this problem under a long-term time-average resource constraint. However, in real systems with instantaneous resource constraints, how to optimally exploit the temporal correlation and satisfy realistic stringent constraint on the instantaneous
service remains elusive. In this work, we incorporate a stringent constraint on the simultaneously scheduled users and propose a low-complexity scheduling algorithm that dynamically implements user scheduling and dummy packet broadcasting. We show that the throughput region of the optimal policy under the long-term average resource constraint can be asymptotically achieved in the stringent constrained scenario by the proposed algorithm, in the many users limiting regime.
\end{abstract}

\section{Introduction}

In wireless networks, the states of the wireless channels fluctuate in time. This characteristic calls for designing resource allocation algorithms that dynamically adapt to the random variation of the wireless channels. Scheduling algorithms are essential components of resource allocation. A scheduling algorithm is designed to control a subset of users to consume the scarce network resources (e.g., bandwidth, power, time), so that the overall network utility (e.g., throughput, fairness) is maximized subject to link interference and queue stability constraints. Under the assumption that accurate instantaneous Channel State Information (CSI) is available at the scheduler, maximum-weight-type scheduling algorithms (e.g., \cite{MWM}-\cite{Eryilmaz05}) are known to be throughput-optimal, i.e., they can maintain system stability for arrival rates that are supportable by any other scheduler.

The performance of efficient scheduling algorithm relies heavily on the accurate instantaneous CSI at the scheduler. In practice, however, accurate instantaneous CSI is difficult to obtain at the scheduler, i.e., a significant amount of system resources must be spent to accurately estimate the instantaneous CSI (see e.g., \cite{Tse}). Therefore, acquiring CSI continuously from all users is resource-consuming and impractical as the size of network increase. Hence, in this work we consider the important scenario where the instantaneous CSI is not directly accessible to the scheduler, but is instead learned at the user and fed back to the scheduler via ARQ-styled feedback after a certain delay. Many scheduling algorithms have been designed that consider imperfect CSI, where the channel state is considered as independent and identically distributed (i.i.d.) processes over time (e.g., \cite{Allerton}-\cite{Wenzhuo_wiopt13}). However, although the \emph{i.i.d.} channel models facilitate trackable analysis, it does not capture the time-correlation of the fading channels.

Because perfect instantaneous CSI is costly to acquire, the time-correlation or channel memory inherent in the fading channels is an important resource that can be exploited by the scheduler to make more informed decisions, and hence to obtain significant throughput/utility gains (e.g., \cite{WiOpt_version}-\cite{CelicModiano}). Under imperfect CSI, channel memory, and resources constraint, the scheduler needs to intelligently balance the intricate `exploitation-exploration tradeoff', i.e., to decide at each slot whether to exploit the channels with more up-to-date CSI, or to explore the channels with outdated CSI.

We consider the downlink of a single cell, where the packets destined to each user are stored in a corresponding data queue for transmission. Under the complicated channel memory evolution and queue evolution, traditional Dynamic Programming based approaches can be used for designing scheduling schemes, but are intractable due to the well-known `curse of dimensionality'. Recently, a low-complexity algorithm was proposed in \cite{WiOpt_version} that considers throughput-optimal downlink scheduling with imperfect CSI over time-correlated fading channels, under a constraint on the \emph{long-term average} number of transmissions.

Scheduling in wireless systems is typically subject to \emph{stringent instantaneous} constraints, such as instantaneous resource limitations from bandwidth, power, interference, etc. In this work, we study scheduling with imperfect CSI over time-correlated channels and under stringent resource constraint where the instantaneous scheduling decision is subject to constraint on the maximum number of scheduled users. The stringent constraint brings with it significant challenges, and to the best of our knowledge under the setting of imperfect CSI, no low-complexity algorithm exists that is optimal for general scenarios.  Under the restrictive regime where users have identical ON/OFF Markovian channel statistics, round-robin based scheduling policies are shown to be throughput optimal in \cite{Neely_RR}\cite{Neely_RR2}. Further, under these settings, it has been shown in \cite{ZhaoTWC}\cite{SM_IT} that greedy scheduling algorithms are also optimal. In \cite{KrishnaModiano}\cite{CelicModiano}, throughput-optimal frame-based policies are proposed. These policies rely on solving a Linear Programming in each frame, which is hindered by the curse of dimensionality where the computational complexity grows exponentially with the network size.

In this paper, we propose a low-complexity algorithm in wireless downlink under stringent constraint and heterogeneous Markovian transition statistics across users. We prove that the proposed algorithm has asymptotical optimal properties in the regime of a large number of users. Our contributions are as follows:

\begin{itemize}

\item Under stringent constraint on the instantaneous number of transmissions, we propose a novel low-complexity \emph{joint scheduling and broadcasting} algorithm. At each slot, the scheduler dynamically decides whether to schedule a subset of users and learn their channel state feedback via ARQ-styled feedback, or to broadcast a dummy packet to a larger set of users to learn their channel states from ARQ-styled feedback but with no throughput gain.

\item We conduct our analysis in the framework of Partially Observable Markov Decision Process, where we utilize Whittle¡¯s index analysis of Restless Multi-armed Bandit Problem (RMBP) \cite{Whittle}. We then use a \emph{Large-Deviation-based Lyapunov} technique over time frames to prove the throughput performance of the proposed algorithm.

\item We prove that, the throughput region in \cite{WiOpt_version}, which is achieved by an optimal policy under a relaxed constraint on the long-term average number of transmissions, can be asymptotically achieved in the stringent constrained scenario by the proposed algorithm, in the regime of a large number of users.

\end{itemize}

\begin{figure}
\centering
\includegraphics[width=2in]{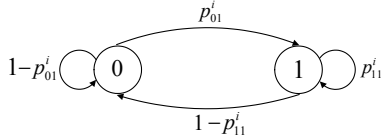}
\caption{Two state Markov Chain model.}
\label{fig:chain}
\end{figure}

\section{System Model}

\subsection{Downlink Scheduling Problem}

We study a wireless downlink network with one Base Station (BS) and $N$ users. Time is slotted with each slot synchronized among BS and users. Each user $i$ occupies a dedicated wireless channel, whose state is denoted by $C_i[t]$ at slot $t$. The channel state $C_i[t]$ evolves as an ON/OFF Markov chain across time slots with state space $\mathcal{S}=\{0, 1\}$, independently of other channels. Channel state `1' represents high channel gain where one packet can be transmitted successfully through the channel, whereas state `0' represents deep fading state where no packet can be delivered\footnote{Our results easily extend to general two-state scenarios where multiple packets, different across channels, can be transmitted in the two states.}. The Markovian channel state evolution is depicted in Fig.~\ref{fig:chain}, represented by the transition probabilities
\begin{align}
p^i_{jk}:= \Pr\big(C_l[t]{=}k\big | C_i[t{-}1]{=}j\big), j,k\in\mathcal{S}.\nonumber
\end{align}

We assume that $p^i_{11} > p^i_{01}$ for $i{=}1,2,\cdots, N$. This assumption implies positive correlation and is commonly made in this field (e.g., \cite{Wenzhuo_infocom12}\cite{Neely_RR}\cite{KrishnaModiano}\cite{sugu_aslm}), which means that auto-correlation of the channel state process is non-negative \cite{Javidi}. We also assume that there exists a positive constant $\delta>0$ so that $p^i_{01}{>}\delta$ and $p^i_{10}{>}\delta$ for all $i$ to allow at least minimum probability of cross transition between the two states, which captures the random varying nature of the wireless channels. Our result, however, can be extended to more general scenarios.

Data packets destined for different users are stored in separate queues at the BS before they are successfully transmitted. The queue length for user $i$ at slot $t$ is denoted by $q_i[t]$. The number of data packet that arrives at queue $i$ for the $i$-th user is denoted as $A_i[t]$, which forms an \emph{i.i.d.} process with mean $\lambda_i$ and a bounded second moment.

At the beginning of every time slot, the scheduler at the BS selects users for data transmission. We let $a_i[t]\in\{0,1\}$ indicate whether user $i$ is scheduled at slot $t$. The $i$-th data queue evolves as $q_i[t{+}1]{=}\max \{ 0, q_i[t] {-}a_i[t]{\cdot} C_i[t] \}{+} A_i[t]$.

Due to the afore-mentioned resource constraints, the scheduling decisions are made without the exact knowledge of the channel state in the current slot. In our model, the scheduler at the BS obtains the accurate CSI via ARQ-styled ACK/NACK feedback, only from the scheduled users \emph{at the end of each slot} following data transmission, i.e., an ACK from scheduled user $i$ implies $C_i[t]=1$, while an NACK implies $C_i[t]=0$.

\begin{figure}
\centering
\includegraphics[width=2.4in]{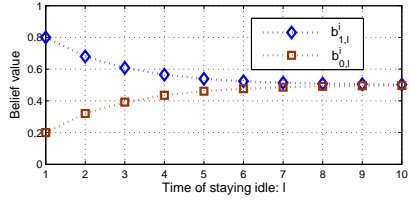}
\caption{Belief value evolution, $p^i_{11}=0.8$, $p^i_{01}=0.2$, $b^i_s=0.5$.}
\label{fig:belief_evol}\vspace{-15pt}
\end{figure}

We consider the class $\Phi$ of (possibly non-stationary) scheduling policies that make scheduling decisions based on the history of observed channel states, arrival processes, and scheduling decisions. Under the aforementioned instantaneous constraint from power, bandwidth and interference, the scheduling schemes are subject to the constraint that the number of scheduled transmissions is under $M$ at each time slot $t$, i.e.,
\begin{align}
\sum_{i=1}^{N} a_i^{\phi}[t] \leq M, \label{eq:constraint}
\end{align}
where $M\leq N$, and $a_i^{\phi}[t]\in\{0,1\}$ indicates if the $i$-th user is scheduled at slot $t$ under policy $\phi \in \Phi$. For example, in wireless cellular downlink, $M$ can correspond to the number of orthogonal time-bandwidth slices, where one user can be scheduled in each slice without causing interference to other users.

\subsection{Belief Value Evolution}

The scheduler maintains a belief value $\pi_i[t]$ for each channel $i$, defined as the probability of channel $i$ being in state $1$ at the beginning of $t$-th slot conditioned on the past channel state observations. The belief values are hence updated according to the scheduling decisions and accurate channel state feedbacks,
\begin{align}
\pi_i[t+1]=
\begin{cases}
p^i_{11}& \text{if $a_i[t]=1$ and $C_i[t]=1$,}\\
p^i_{01}& \text{if $a_i[t]=1$ and $C_i[t]=0$,} \\
Q_i(\pi_i[t])& \text{if $a_i[t]=0$,}
\end{cases} \label{eq:evolve}
\end{align}
where $Q_i(x){=}x p^i_{11}+(1{-}x)p^i_{01}$ is the belief evolution operator when user $i$ is not scheduled in the current slot. In our setup, the belief values are known to be sufficient statistics to represent the past scheduling decisions and channel state feedback \cite{Sondik_thesis}. In the meanwhile, the belief value $\pi_i[t]$ is the expected throughput for user $i$ if it is scheduled in slot $t$.

For the $i$-th user, we use $b^i_{c,l}$ to denote the state of its belief value when the most recent channel state was observed $l$ time slots ago and was in state $c\in \{0,1\}$. The closed form expression of $b^i_{c,l}$ can be calculated from (\ref{eq:evolve}) and is given as
\begin{align}
b^i_{0,l}{=}\frac{p^i_{01}\hspace{1pt}{-}\hspace{1pt}(p^i_{11}{-}p^i_{01})^l
p^i_{01}}{1+p^i_{01}-p^i_{11}},
b^i_{1,l}{=}\frac{p^i_{01}{+}(1{-}p^i_{11})(p^i_{11}{-}p^i_{01})^l}{1+p^i_{01}-p^i_{11}}.
\nonumber
\end{align}

As depicted in Fig.~\ref{fig:belief_evol}, if the scheduler is never informed of the $i$-th user's channel state, the belief value monotonically converges to the stationary probability $b_s^{i}{:=}{p^i_{01}}/{(1+p^i_{01}-p^i_{11})}$ of the channel being in state $1$. We assume that the belief values of all channels are initially set to their stationary values. It is then clear that, based on (\ref{eq:evolve}), each belief value $\pi_i[t]$ evolves over a countable state space, denoted by $\mathcal{B}_i{=}\{b^i_{s}, b^i_{c,l}: c \hspace{1pt}{\in}\hspace{1pt} \{0,1\}, l\hspace{1pt}{\in}\hspace{1pt} \mathbb{Z}^+ \}$.

\subsection{Network Stability Regions}

We adopt the following definition of queue stability
\cite{MWM}: queue $i$ is stable if there exists a limiting
stationary distribution $F_i$ such that $\lim_{t\rightarrow \infty}
P( q_i[t] \leq q)= F_i(q)$.  When there are $N$ total downlink users and at most $M$ users can be simultaneously scheduled, the \emph{network stability region} $\bm \Lambda_{str}^{N,M}$ is defined as the closure of the set of arrival rate vectors supported by all policies in class $\Phi$ that does not lead to system instability while abiding by the stringent constraint (\ref{eq:constraint}).

For comparison purpose, we introduce another region $\bm \Lambda_{rel}^{N,M}$ as the closure of the set of arrival rate vectors supported by all policies in class $\Phi$ that maintains queue stability and satisfies the following \emph{relaxed constraint} that only requires an \emph{average number} of $M$ users to be activated in the long run,
\begin{align}
\limsup_{T\rightarrow \infty} \frac{1}{T}\mathbb{E}\Big[\sum_{t=0}^{T-1} \sum_{i=1}^{N} a_i^{\phi}[t]\Big ]\leq M.\label{eq:relaxed_cons}
\end{align}

The region $\bm \Lambda_{rel}^{N,M}$ provides a benchmark for our analysis on the scenarios with stringent constraint. Note that, contrary to the stringent constraint (\ref{eq:constraint}), the relaxed constraint (\ref{eq:relaxed_cons}) allows the activation of more than $M$ users in each time slot, provided the long term average number of transmissions does not exceed $M$. Hence the corresponding region $\bm \Lambda_{rel}^{N,M}$ provides an upper bound to the region $\bm \Lambda_{str}^{N,M}$ under the stringent constraint. In the paper propose a policy that not only abides by the stringent constraint, but also asymptotically achieves the stability region upper bound $\bm \Lambda_{rel}^{N,M}$.

\section{Optimal Policy for Weighted Sum-throughput Maximization under a Relaxed Constraint}
\label{sec:sum_thr_rel}

We begin our analysis by introducing an optimal algorithm for weighted sum-throughput maximization under the associated relaxed constraint. The corresponding algorithm serves as an essential part in our main result.

Specifically, consider the following weighted sum-throughput maximization problem $\Psi_{rel}(\mathbf{r}, N, M)$ for a given vector $\mathbf{r}=(r_i)_{i=1}^N$, where the expected service rate for each user $i$ is scaled by a non-negative factor $r_i$,
\begin{align}
\max_{\phi \in \Phi}& \ \liminf_{T\rightarrow \infty} \frac{1}{T} \mathbb{E}\Big[\sum_{t=0}^{T-1} \sum_{i=1}^N r_i {\cdot} \pi_i[t] {\cdot} a_i^{\phi}[t] \Big]\label{eq:obj_rel}\\
\text{s.t.}&\hspace{5pt} \limsup_{T\rightarrow \infty} \frac{1}{T}\mathbb{E}\Big[\sum_{t=0}^{T-1} \sum_{i=1}^{N} a_i^{\phi}[t]\Big ]\leq M. \label{eq:cons_rel}
\end{align}

The above problem $\Psi_{rel}(\mathbf{r}, N, M)$ is hence a constrained Partially Observable Markov Decision Process (CPOMDP) \cite{Eitan}\cite{Meyn_CPOMDP}.

The problem~(\ref{eq:obj_rel})-(\ref{eq:cons_rel}) can be tackled in the framework of the Restless Multiarmed Bandit Problem (RMBP) \cite{Whittle} by making use of the associated Whittle's indexability analysis. In the rest of this section, we give a brief review of the Whittle's indices for RMBP \cite{Whittle}\cite{Zhao_index}, and the optimal algorithms proposed in \cite{WiOpt_version} for solving this problem. For details of our description, please refer to \cite{WiOpt_version}\cite{WiOpt_TONversion}\cite{Zhao_index}\cite{Whittle}.

\subsection{Whittle's Index for Restless Multi-armed Bandit Problem}

RMBPs refer to a collection of sequential dynamic resource allocation problems where several independently evolving projects compete for service. In each slot, a subset of these competing projects is served. The state of each project stochastically evolves over time, based on the current state of the project and on whether the project is served in the slot. Serving a project brings a reward whose value depends on its state. Hence, in RMBPs, the controller needs to consider the fundamental tradeoff between decisions that bring high instantaneous rewards, versus those decisions that bring better future rewards but sacrifices the instantaneous rewards. Solving RMBPs are known to be hard.

Whittle's index analysis \cite{Whittle} for RMBPs considers the following \emph{virtual system}: in each slot, the controller makes one of the two decisions for each project $P$: (1) Serve project $P$ and accrue an immediate reward as a function of its state which is the same as in the original RMBP. (2) Do not serve project $P$ and obtain an immediate reward $\omega$ for passivity. The state evolution of the project $P$ is the same as in the original RMBP, depending on its current state and current action. In this virtual system, the design goal is to maximize the long-term expected reward by balancing the `reward for serving' and the `subsidy for passivity' in each slot.

Letting $\mathcal{I}(\omega)$ denote the set of states of project $P$ in which the optimal action is to stay passive, the Whittle's indexability condition is defined as follows.

\textit{Project $P$ is Whittle indexable if the set $\mathcal{I}(\omega)$ monotonically increases from $\emptyset$ to the state space $\mathcal{S}$ of project $P$, as $\omega$ increases from $-\infty$ to $\infty$. The RMBP is Whittle indexable if \textit{every} project is Whittle indexable.}

If Indexability holds, for each state $s$ of a project, the \emph{Whittle's index} $W(s)$ is defined as the infimum of $\omega$ in which it is optimal to stay idle in the $\omega$-subsidized system, i.e.,
\begin{align}
W(s)=\inf\{ \omega: s \in \mathcal{I}(\omega) \}. \nonumber
\end{align}

Under an average constraint on the number of projects scheduled per slot, it is known that, upon the existence of the Indexability condition, a low-complexity algorithm exists based on the `Whittle's indices': activate the projects with large Whittle's index value \cite{Whittle}.

The RMBP theories and the associated Whittle's indices can be used in our downlink scheduling problem. Here, each downlink user corresponds to a project in the RMBP, with the associated state being the belief value of its channel. Correspondingly, the project is considered served if the user is scheduled for data transmission at a slot. Hence the Whittle's index policy, because of its simplicity, is very attractive to provide optimal yet low-complexity solutions problem $\Psi_{rel}(\mathbf{r}, N, M)$.

\subsection{Optimal Policy for Weighted Sum-throughput Maximization under a Relaxed Constraint}

It was shown in that our downlink scheduling problem is Whittle indexable \cite{Zhao_index}, and, under uniform weight vector $\mathbf{r}{=}\mathbf{1}$, an optimal policy for problem $\Psi_{rel}(\mathbf{1}, N, M)$ exists based on Whittle's indexability analysis of Restless Multi-armed Bandit Problem \cite{Wenzhuo_infocom12}. Specifically, for channel $i$, a closed form \emph{Whittle's index value} $W_i^{\mathbf{1}}(\pi)$ is assigned to each belief state $\pi \in \mathcal{B}_i$. These indices intelligently capture the exploitation-exploration value to be gained from scheduling the user at the corresponding belief state \cite{Wenzhuo_infocom12}. The closed form expression of the Whittle's index value $W_i^{\mathbf{1}}(\pi), \pi \in \mathcal{B}_i$, is given as follows \cite{Wenzhuo_infocom12}\cite{Zhao_index},
\begin{align}
\label{eq:indices}
W_i^{\mathbf{1}}(\pi)
{=}\hspace{-3pt}\begin{cases}
\frac{(\pi-Q_i(\pi)) (l+1)+Q_i(\pi)}{1-p^i_{11}+(\pi-Q_i(\pi))l+Q_i(\pi)} &\text{if $p^i_{01} {\leq} \pi{=}b^i_{0,l} {<} b^i_s$} \\
\frac{p^i_{01}}{(1-p^i_{11})(1+p^i_{01}-p^i_{11})+p^i_{11}} &\text{if $b^i_s \leq \pi\leq p^i_{11}$}
\end{cases}
\end{align}

It was shown that $W_i^{\mathbf{1}}(\pi)$ monotonically increases with $\pi$ and satisfies $W_i^{\mathbf{1}}(\pi) \in [0,1]$ \cite{Wenzhuo_infocom12}\cite{Zhao_index}. The following lemma gives an optimal algorithm to the problem $\Psi_{rel}({\mathbf{r}}, N, M)$ with arbitrary non-negative weight vector $\mathbf{r}$. The proof of the lemma can be found in \cite{WiOpt_version}\cite{Wenzhuo_infocom12}.

\begin{lemma}
\label{lemma:thres_index}
There exists an optimal policy $\phi_{rel}^*(\mathbf{r},N,M)$ for problem $\Psi_{rel}(\mathbf{r}, N, M)$ (cf. (\ref{eq:obj_rel})-(\ref{eq:cons_rel})), parameterized by a threshold $\omega^*$ and a randomization factor $\rho^*$, such that

\noindent(i) The scheduler maintains an $\mathbf{r}$-weighted index value $W^{\mathbf{r}}_i(\pi_i[t])=r_i \cdot W_i^{\mathbf{1}}(\pi_i[t])$ for user $i$.

\noindent(ii) User $i$ is scheduled if $W^{\mathbf{r}}_i(\pi_i[t])\hspace{1pt}{>}\hspace{1pt}\omega^*$, and stays idle if \hspace{3pt}$W^{\mathbf{r}}_i(\pi_i[t])\hspace{1pt}{<}\hspace{1pt}\omega^*$.
If $W^{\mathbf{r}}_i(\pi_i[t])\hspace{1pt}{=}\hspace{1pt}\omega^*$, it is
scheduled with probability $\rho^*$.

\noindent(iii) The parameters $\omega^*$ and $\rho^*$ are such that the long-term average number of transmissions equals $M$.
\end{lemma}

\subsection{Approximate $\omega^*$ and $\rho^*$ by $\omega_{\tau}$ and $\rho_{\tau}$}

Note that the parameters $\omega^*$ and $\rho^*$ need to be carefully chosen to satisfy the complementary slackness condition, i.e., Lemma~\ref{lemma:thres_index}(iii). While directly finding $\omega^*$ and $\rho^*$ may be difficult, an algorithm was proposed in \cite{WiOpt_version} to derive approximate values of $\omega^*$ and $\rho^*$ based on a fictitious model over \emph{truncated belief state space}. Over the fictitious model, the belief value of a user is set to its steady state if the corresponding channel has not been scheduled for $\tau$ slots. Specifically, the algorithm $G^{\tau}(\mathbf{r},N, M)$ was introduced \cite{WiOpt_version}\cite{WiOpt_TONversion} to calculate $\omega_{\tau}$ and $\rho_{\tau}$.

\renewcommand{\thealgorithm}{}

\algnewcommand{\algorithmicgoto}{\textbf{go to}}%
\algnewcommand{\Goto}[1]{\algorithmicgoto~\ref{#1}}%

\begin{algorithm}
  \caption{\hspace{-6pt}$G^{\tau}(\mathbf{r},N, M)$\textbf{:} Calculation of $\omega_{\tau}$ and $\rho_{\tau}$}\label{alg:thres}
  \begin{algorithmic}[1]
  \State $\text{TxTime}[i]=1 \text{ for all $i\in\{1,\cdots, N\}$}$
  \State $\text{TotalTime}=N$
  \State \text{\textbf{struct} Index}
  \State $\{$\text{ float value}
  \State \text{ \hspace{3pt} int user}
  \State $\}$ $\mathbf{I}[(2\tau+1)N],\mathbf{w}[(2\tau+1)N]$
  \vspace{5pt}\State $j=0$
  \For{$i=1$ \text{to} $N$}
  \For{each $\pi_i \in \mathcal{B}_i^{\tau}$}
        \State $W_i^{\mathbf{r}}(\pi_i)= r_i\cdot W_i^{\mathbf{1}}(\pi_i)$
    \State $\mathbf{I}[j].$\text{value}$=W_i^{\mathbf{r}}(\pi_i)$
    \State $\mathbf{I}[j].$\text{user}$=i$
    \State $j\gets j+1$
  \EndFor
  \EndFor
  \vspace{5pt}\State $\mathbf{w}=$\text{sort}$(\mathbf{I})$
    \For{$k=1$ \text{to} \text{size}$(\mathbf{w})$}
    \vspace{5pt}\State $\text{NewTime}[\mathbf{w}[k].\text{user}]={\alpha}^{\tau}_{\mathbf{w}[k].\text{user}}(\mathbf{w}[k].\text{value}, 1)$
\State  $\text{TimeDiff}=\text{TxTime}[\mathbf{w}[k].\text{user}]{-}\text{NewTime}[\mathbf{w}[k].\text{user}]$
    \State $\text{TotalTime}=\text{TotalTime}-\text{TimeDiff}$
    \If{$\text{TotalTime}<M$}
    \State $\omega_{\tau}=\mathbf{w}[k{-}1].\text{value}$
    \State $\text{TxTime}[\mathbf{w}[k{-}1].\text{user}]=M{-}\hspace{-9pt}\sum\limits_{i\neq \mathbf{w}[k{-}1].\text{user}}\hspace{-9pt}\text{TxTime}[i]$
    \State $\rho_{\tau}=\beta_{\mathbf{w}[k{-}1].\text{user}}(\omega_{\tau},\text{TxTime}[\mathbf{w}[k{-}1].\text{user}])$
    \State \text{\textbf{Break}}
    \EndIf
    \State $\text{TxTime}[\mathbf{w}[k].\text{user}]{=}\text{NewTime}[\mathbf{w}[k].\text{user}]$
    \EndFor
    \State \textbf{return} $\omega_{\tau}$, $\rho_{\tau}$
\end{algorithmic}
\end{algorithm}

\subsection{Policy with approximate parameters $\omega_{\tau},\rho_{\tau}$}
\label{sec:thr_appro_alg}

The next policy, denoted as $\phi_{rel}^{\tau}(\mathbf{r},N,M)$, uses the approximated parameters $\omega_{\tau}$ and $\rho_{\tau}$ over the \emph{original untruncated model}.

\begin{algorithm}
  \caption{$\phi_{rel}^{\tau}(\mathbf{r},N,M)$\textbf{:} $\mathbf{r}$-weighted Index Policy}\label{alg:thres}
  \begin{algorithmic}[1]
  \State \textbf{Initialization phase:} The parameters $\omega_{\tau}$ and $\rho_{\tau}$ are calculated by algorithm $G^{\tau}(\mathbf{r},N,M)$.
  \State \textbf{At slot $\bm t$:} user $i$ is scheduled if the $\mathbf{r}$-weighted index value $W^{\mathbf{r}}_i(\pi_i[t])> \omega_{\tau}$, and stays passive if \hspace{3pt}$W^{\mathbf{r}}_i(\pi_i[t])<\omega_{\tau}$.
If $W^{\mathbf{r}}_i(\pi_i[t])=\omega_{\tau}$, user $i$ is
scheduled with probability $\rho_{\tau}$.
\end{algorithmic}
\end{algorithm}

\noindent\textbf{Remark:} The computational complexity of the initialization phase of algorithm $\phi_{rel}^{\tau}(\mathbf{r},N,M)$ is dominated by sorting the index values in Algorithm $G^{\tau}(\mathbf{r},N,M)$ (line 16), which has complexity $O\big((2\tau+1)N \cdot \log \big((2\tau+1)N\big)\big)$.

We let $V^*(\mathbf{r},N,M)$ be the weighted sum-throughput under the optimal policy $\phi_{rel}^*(\mathbf{r},N,M)$, and let $V_{\tau}(\mathbf{r},N,M)$ be that under policy $\phi_{rel}^{\tau}(\mathbf{r},N,M)$, i.e.,
\begin{align}
&V^*_{rel}(\mathbf{r},N,M)\nonumber\\
=&\liminf_{T\rightarrow \infty} \frac{1}{T} \mathbb{E}\Big[\sum_{t=0}^{T-1} \sum_{i=1}^N r_i {\cdot} \pi_i[t] {\cdot} a_i^{\phi_{rel}^*(\mathbf{r},N,M)}[t] \Big],\label{eq:thr_nontrun}
\end{align}
\begin{align}
&V_{rel}^{\tau}(\mathbf{r},N,M)\nonumber\\
=&\liminf_{T\rightarrow \infty} \frac{1}{T} \mathbb{E}\Big[\sum_{t=0}^{T-1} \sum_{i=1}^N r_i {\cdot} \pi_i[t] {\cdot} a_i^{\phi_{rel}^{\tau}(\mathbf{r},N,M)}[t] \Big].\label{eq:thr_trun}
\end{align}

The policy $\phi_{rel}^{\tau}(\mathbf{r},N,M)$ provides throughput arbitrarily close to $V_{rel}^*(\mathbf{r},N,M)$ as the truncation size increases, while abiding the long-term average number of transmissions constraint, which was shown in \cite{WiOpt_version}\cite{WiOpt_TONversion} and recorded below.

\begin{lemma}\label{lemma:eps_bound_tau}
\noindent For $\tau\geq\tau_0:=\Big\lceil4 \max\big\{\frac{1}{{-}\log(2\delta)}, \frac{1}{\log^2(2\delta)}\big\}\Big\rceil$,

\noindent(i) The throughput performance difference between the policies $\phi_{rel}^*(\mathbf{r},N,M)$ and $\phi_{rel}^{\tau}(\mathbf{r},N,M)$ is bounded by
\begin{align}
|V_{rel}^*(\mathbf{r},N,M)-V_{rel}^{\tau}(\mathbf{r},N,M)| \leq f(\tau) \sum_{i=1}^N r_i,
\end{align}

\noindent where $f(\tau){=}\sum_{i=1}^N f_i(\tau)$, which satisfies $f(\tau){\rightarrow} 0$ as $\tau{\rightarrow}\infty$ with
\begin{align}
\label{eq:act_time0}
f_i(\tau)=\frac{1+b^i_{0,\tau}-p^i_{11}}{b^i_{0,\tau}{+}(1{-}p^i_{11})\cdot\tau}.
\end{align}

\noindent(ii) The long-term average number of transmissions under policy $\phi_{rel}^{\tau}(\mathbf{r}, N, M)$ satisfies the relaxed constraint~(\ref{eq:cons_rel}).
\end{lemma}

\section{Weighted Sum-throughput Maximization Problem under Stringent Constraint}

Note that, although the algorithm $\phi_{rel}^{\tau}\big(\mathbf{r}, N, M\big)$ in last section abides by the relaxed long term average constraint~(\ref{eq:cons_rel}) on the number of users scheduled, the number of users scheduled in \emph{each instantaneous slot} can violate the stringent interference constraint~(\ref{eq:constraint}) that requires no more than $M$ users scheduled at a slot. Hence the corresponding stability region $\bm \Lambda_{rel}^{N,M}$ provides an upper bound on $\bm\Lambda_{str}^{N,M}$.

In this section, we also consider the $\mathbf{r}$-weighted sum throughput optimization problem as in the last section where the throughput of user $i$ is scaled by a factor $r_i$, but under the stringent constraint, i.e., no more than $M$ users are scheduled for data transmission at each time slot. we propose a joint scheduling and broadcasting algorithm that leverages the policy in the previous section for the stringent constrained problem. This algorithm has novelty of incorporating the possibility of \emph{broadcasting} a dummy packet at a slot, and can provide performance asymptotically close to algorithm $\phi_{rel}^{\tau}\big(\mathbf{r}, N, M\big)$ for the relaxed problem in the regime of large values of $N$.

\subsection{Policy with Joint Scheduling and Broadcasting}

The proposed policy, denoted by $\phi_{str}^{\tau}\big(\mathbf{r}, N, M, K\big)$ with $K\leq M$, builds on the policy $\phi_{rel}^{\tau}\big(\mathbf{r}, N, M\big)$ for the relaxed problem. However, it fundamentally differs from $\phi_{rel}^{\tau}\big(\mathbf{r}, N, M\big)$ in the following way. At the beginning of each slot, algorithm $\phi_{str}^{\tau}\big(\mathbf{r}, N, M, K\big)$ carefully makes one of two choices: 1) transmit data packets to no more than $M$ users and receive ARQ-type feedback from them, or 2) \emph{broadcast} a dummy packet to more than $M$ users, and learn their channel states from their ARQ-type feedback. Note that, the dummy packet is known to the users and contains no new information and hence does not bring throughput gains if it is broadcasted. However, the scheduler still receive ARQ-styled feedback from the candidates, and hence obtain CSI update from possibly more than $M$ users.

The parameter $K$ controls how aggressively the dummy packets are broadcasted. As we will see next, intelligently tuning this parameter is important for the asymptotic optimality result of the proposed algorithm.

Recall that the $\mathbf{r}$-weighted index value is defined in Lemma~\ref{lemma:thres_index}. Algorithm $\phi_{str}^{\tau}\big(\mathbf{r}, N, M, K\big)$ is proposed next.

\begin{algorithm}
  \caption{\hspace{-4pt}$\phi_{str}^{\tau}\big(\mathbf{r}, N, M, K\big)$\textbf{ under stringent constraint}}\label{alg:thres}
  \begin{algorithmic}[1]
  \State \textbf{Initialization phase:} The parameters $\omega_{\tau}$ and $\rho_{\tau}$ are calculated by algorithm $G^{\tau}(\mathbf{r},N,K)$.
  \State \textbf{At slot $\bm t$, candidate selection:} user $i$ is called a `\emph{candidate}', represented by $\theta_i[t]{=}1$, if the $\mathbf{r}$-weighted index value $W^{\mathbf{r}}_i(\pi_i[t]){>}\omega_{\tau}$, and is not a candidate, i.e., $\theta_i[t]{=}0$, if \hspace{3pt}$W_i^{\mathbf{r}}(\pi_i[t])<\omega_{\tau}$.
If $W_i^{\mathbf{r}}(\pi_i[t])=\omega_{\tau}$, user $i$ becomes a `candidate' with probability $\rho_{\tau}$.
  \State \textbf{At slot $\bm t$, transmission:} If the total number of candidates is under $M$, i.e., $\sum_{i=1}^N \theta_i[t] \leq M$, then all the candidates are scheduled for data transmission, i.e., $a_i^{\phi_{str}^{\tau}\big(\mathbf{r}, N, M, K\big)}[t]=\theta_i[t]$ for all $i$. If there are more than $M$ candidates, then $a_i^{\phi_{str}^{\tau}\big(\mathbf{r}, N, M, K\big)}[t]=0$ for all $i$, and dummy packet is \emph{broadcasted}.
  \State \textbf{At slot $\bm t$, feedback:} At the end of each slot, if data packets are transmitted, the scheduled users send ARQ feedback to the BS; if the dummy packet is broadcasted, the \emph{candidates} send ARQ feedback to the BS. The belief values are updated based on the feedback.
\end{algorithmic}
\end{algorithm}

\newpage
We next give a step-by-step explanation of this algorithm.

\vspace{10pt}\noindent\textbf{Remarks:}

\noindent(1) Steps 1-2 of algorithm $\phi_{str}^{\tau}\big(\mathbf{r}, N, M, K\big)$ is exactly algorithm $\phi_{rel}^{\tau}\big(\mathbf{r}, N, K\big)$, where the scheduled users in algorithm $\phi_{rel}^{\tau}\big(\mathbf{r}, N, K\big)$ becomes the candidates in $\phi_{str}^{\tau}\big(\mathbf{r}, N, M, K\big)$.

\noindent (2) Step 3 ensures that the stringent interference constraint is met so that data packets are transmitted to no more than $M$ users. Hence if the number of candidates exceeds $M$, a dummy packet is broadcasted for the scheduler to learn the channel states of the candidates and no throughput is accrued.

\noindent (3) Because of step 4, the scheduler receives channel state feedback from all the candidates, although data packets may not be transmitted. By taking this approach, the channel memory evolution in the relaxed constrained algorithm $\phi_{rel}^{\tau}\big(\mathbf{r}, N, K\big)$ is maintained in the stringent constrained algorithm $\phi_{str}^{\tau}\big(\mathbf{r}, N, M, K\big)$, which facilities much more trackable performance analysis.

\noindent (4) In step 4, only the candidates (instead of all users) send feedback to the BS if dummy packet is broadcasted. By allowing only the candidates to feedback\footnote{This can be achieved by marking the corresponding bits in the dummy packet.}, the algorithm not only helps maintain the tractability of channel memory evolution, more importantly, it fits with the realistic scenario where it is costly (in terms of time, power, bandwidth, etc.) to obtain feedback from all users, especially when user number is large.

We henceforth let $V_{str}^{\tau}\big(\mathbf{r}, N, M, K\big)$ be the weighted sum-throughput under policy $\phi_{str}^{\tau}\big(\mathbf{r}, N, M, K\big)$, i.e.,
\begin{align}
&V_{str}^{\tau}\big(\mathbf{r}, N, M, K\big)\nonumber\\
{=}&\liminf_{T\rightarrow \infty} \frac{1}{T} \mathbb{E}\Big[\sum_{t=0}^{T-1} \sum_{i=1}^N r_i {\cdot} \pi_i[t] {\cdot} a_i^{\phi_{str}^{\tau}\big(\mathbf{r}, N, M, K\big)}[t] \Big].\label{eq:thr_tru_str}
\end{align}

\subsection{Performance of the algorithm under stringent constraint}

From the algorithm and Remark (1) thereafter, in each slot, if the number of scheduled users exceeds $M$ under algorithm $\phi_{rel}^{\tau}\big(\mathbf{r}, N, K\big)$ for the relaxed problem, the number of candidates under algorithm $\phi_{str}^{\tau}\big(\mathbf{r}, N, M, K\big)$ exceeds $M$ and a dummy packet is broadcasted, otherwise all candidates are scheduled for data transmission. Hence in the regime when $K$ is close to $M$, the larger the $K$, the more aggressively are dummy packets broadcasted, which bring more updated system-level channel state information, but with a tradeoff that no throughput is obtained in these broadcasting slots. On the other hand, in the regime when $K$ is away from $M$, the smaller the $K$, on average there are less candidates and hence scheduled users, which also brings down the throughput.

The next lemma bounds the difference between the throughput performance $V_{str}^{\tau}\big(\mathbf{r}, N, M, K\big)$ of algorithm $\phi_{str}^{\tau}\big(\mathbf{r}, N, M, K\big)$ for the stringent constrained problem, and the throughput $V_{rel}^{\tau}\big(\mathbf{r}, N, M\big)$ of $\phi_{rel}^{\tau}\big(\mathbf{r}, N, M\big)$ for the problem under relaxed constraint. Recall that $V_{str}^{\tau}\big(\mathbf{r}, N, M, K\big)$ and $V_{rel}^{\tau}\big(\mathbf{r}, N, M\big)$ were defined in (\ref{eq:thr_trun}) and (\ref{eq:thr_tru_str}), and $\delta$ was defined in the introduction so that $p^i_{01}>\delta$ and $p^i_{10}>\delta$ for all $i$.

\begin{lemma}
\label{lemma:weighted_sum_conv}
{If $K>M/2$, }then the following bounds hold for the values of $V_{str}^{\tau}(\mathbf{r},N,M,K)$ and $V_{rel}^{\tau}\big(\mathbf{r},N,M\big)$,
\begin{align}
{\mu(M,K)}\leq\frac{V_{str}^{\tau}(\mathbf{r},N,M,K)}{ V_{rel}^{\tau}\big(\mathbf{r},N,M\big)}\leq 1, \label{eq:str_rel_bound}
\end{align}
where
\begin{align}
{\mu(M,K)}{=}\Big[1{-}\exp(-\frac{(M{-}K)^2}{{3K}})\Big]{\cdot}\Big[1{-}\frac{M{-}K}{\delta (K{-}1)}\Big]^+,\label{eq:mu}
\end{align}
and $[\cdot]^+$ represents $\max\{0,\cdot\}$.
\end{lemma}

\noindent \textbf{Proof:} In the proof, we first bound the steady state probability that dummy packets are transmitted using Large Deviation techniques, from which we obtain the first multiplicand in (\ref{eq:mu}). We next bound the effect of $K$ in the throughput different between $V_{str}^{\tau}(\mathbf{r},N,M,K)$ and $V_{rel}^{\tau}\big(\mathbf{r},N,M\big)$, which brings us the second multiplicand in (\ref{eq:mu}). Details of the proof can be found in Appendix~\ref{sec:sum_conv_proof}.\hfill $\blacksquare$

The previous lemma is important to derive the asymptotic throughput performance of the stringent constrained policy, captured in the next proposition. The proposition shows that as both $N$ and $M$ become large, if the parameter $K$ is kept an \emph{appropriate distance} $g(M)$ from $M$, then the throughput performance of policy $\phi_{str}^{\tau}\big(\mathbf{r}, N, M, K\big)$ becomes asymptotically close to $\phi_{rel}^{\tau}\big(\mathbf{r}, N, M\big)$ of the relaxed policy.

\begin{proposition}\label{prop:sum_thr_opt}
Suppose $K=M-{g(M)}$ when $M$ of them can be simultaneously scheduled, where $g(M)\geq 0$ is a function of $M$.

If $g(M)$ satisfies $\lim_{M\rightarrow\infty}g(M)/M=0, \lim_{M\rightarrow\infty} g^2(M)/{M}{=}\infty$, the throughput performance of policy $\phi_{str}^{\tau}\big(\mathbf{r},N,M,M-g(M)\big)$ is asymptotically close to that of $\phi_{rel}^{\tau}\big(\mathbf{r},N,M\big)$, i.e.,
\begin{align}
\label{eq:asymp_opt}\lim_{M\rightarrow \infty}\frac{V_{str}^{\tau}\big(\mathbf{r},N,{M},M{-}g(M)\big)}{V_{rel}^{\tau}\big(\mathbf{r},N,M\big)}=1.
\end{align}
\end{proposition}

\noindent \textbf{Proof:} Since $K=M-g(M)$, from (\ref{eq:str_rel_bound})-(\ref{eq:mu}) we have, if $k>M/2$,
\begin{align}
1\geq&\frac{V_{str}^{\tau}(\mathbf{r},N,M,M-g(M))}{ V_{rel}^{\tau}\big(\mathbf{r},N,M\big)}\nonumber\\
\geq&\mu(M,M-g(M))\nonumber\\
=&\Big[1{-}\exp({-}\frac{g^2(M)}{3\big(M-g(M)\big)})\Big]\Big[1{-}\frac{g(M)}{b^i_s (M{-}g(M){-}1)}\Big]^{+}.\label{eq:conditional_bd}
\end{align}

Since $\lim_{M\rightarrow \infty}g(M)/M=0$ and $\lim_{M\rightarrow\infty} g^2(M)/M=\infty$, we have
\begin{align}
&\lim_{\substack{M\rightarrow \infty}}\hspace{-3pt}\Big[1{-}\exp({-}\frac{g^2(M)}{3\big(M-g(M)\big)})\Big]\hspace{-3pt}\Big[1{-}\frac{g(M)}{b^i_s (M{-}g(M){-}1)}\Big]^{+}=1.\label{eq:conditional_rhs}
\end{align}

Since $\lim_{M\rightarrow \infty}g(M)/M=0$ and $\lim_{M\rightarrow\infty} g^2(M)/M=\infty$, we also have $\lim_{M\rightarrow\infty}K/M=\lim_{M\rightarrow\infty} \big(M-g(M)\big)/M=1>1/2$. Hence from (\ref{eq:conditional_bd})-(\ref{eq:conditional_rhs}), the proposition holds.\hfill $\blacksquare$

\vspace{10pt}\noindent\textbf{Remark:} Proposition~\ref{prop:sum_thr_opt} states that, if the distance between $K$ and $M$ grows at an order larger than $O(\sqrt{{M}})$ but lower than $O(M)$, the performance of the proposed algorithm $\phi_{str}^{\tau}(\mathbf{r},N,M,M-g(N))$ is asymptotically close to $\phi_{rel}^{\tau}(\mathbf{r},N,M)$, which is optimal for the relaxed problem. This is an interesting finding, as it quantities the trade-off between scheduling data packets and broadcasting of dummy packets. When $K$ is less than $O(\sqrt{{M}})$ to $M$, excessive training leaves insufficient slots for data transmission. If $K$ is more than $O(M)$ from $M$, the scheduler is over-conservative on data transmission, which in turn reduces the throughput.

\vspace{-2pt}\section{Queue-based Joint Scheduling and Broadcasting Policy over Time Frames}
%\label{sec:QWI_policy}

Note that, in the two last sections, we considered weighted sum-throughput. In this section, we consider the system model with data queues where \emph{queue stability} is taking into account. In the presence of queue evolution, the problem get much more complicated. Note that, in the sum-throughput optimization problem, the reward of scheduling a user is captured by the Whittle's index value. Under the additional consideration of queue stability, the queue lengths need to be jointly taken into account for scheduling, i.e., a user is scheduled for transmission not only because it has a high index value, but may also because of it has a large queue lengths.

\begin{figure}
\centering
\includegraphics[width=2.2in]{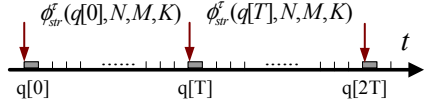}
\vspace{-5pt}
\caption{Illustration of algorithm $\text{Frame}_{\tau}\hspace{-1pt}(T,N,M,K)$.}
\label{fig:Q_frame}\vspace{-10pt}
\end{figure}

In our setup, a simple max-weight-type scheduler (i.e., schedule the $M$ users with the highest $q_i[t] \cdot \pi_i[t]$) can be used, but is no longer optimal. This is because it only exploits the channel condition in the instantaneous slot, i.e., $\pi_i[t]$, but will lose performance since it does not consider exploring outdated channels. Another heuristic scheme is to schedule the $M$ users with the highest multiplication of instantaneous queue length and Index value $q_i[t] \cdot W_i^{\mathbf{1}}(\pi_i[t])$ at each slot $t$. However, it is hard to provide a performance guarantee for this policy, mainly because the Whittle's indexability analysis, which does not consider queue evolution, breaks down if the Whittle's indices are multiplied by queueing length at each instantaneous slot.

Next, we propose a joint scheduling and broadcasting algorithm based on the algorithm $V_{str}^{\tau}(\mathbf{r},N,M,K)$ in the last section. The policy is implemented over separate time-frames and has low-complexity.

We divide the time slots $\{0,1,2,\cdots \}$ into separate \emph{time frames} of length $T$, i.e., the $k$-th frame,  $k\in \{0,1,2,\cdots\}$, includes time slots $kT, {\cdots}, (k+1)T{-}1$. The scheduling decisions in the $k$-th frame are made based on the queue length information $\mathbf{q}[kT]$ at the beginning of that frame. During the $k$-th frame, the policy $\phi_{str}^{\tau}(\mathbf{q}[kT],N,M, K)$, developed in the last section, is implemented. This algorithm is illustrated in Fig.~\ref{fig:Q_frame}. Formally, with N users in the network and under stringent $M$ constraint, the $T$-frame queue-based policy $\text{Frame}_{\tau}\text{(T,N,M,K)}$ is introduced next.

\vspace{-3pt}\begin{algorithm}[H]
  \caption{\hspace{-5pt}$\text{Frame}_{\tau}\hspace{-1pt}(T,N,M,K)$\textbf{:} \hspace{-2pt}$T$-Frame \hspace{-1pt}Queue-based \hspace{-1pt}Policy}\label{alg:thres}
  \begin{algorithmic}[1]
  \vspace{4pt}\State The time slots are divided into frames of length $T$. Slot $t$ is in the $k$-th frame if $kT\leq t < (k+1)T$, $k\in\{0, 1, \cdots\}$.
  \State \textbf{At the beginning of the $\bm k$-th frame:} At the beginning of slot $kT$, implement the algorithm $G^{\tau}(\mathbf{q}[kT],N,K)$ that outputs $\omega_{\tau}$ and $\rho_{\tau}$ for the frame.
  \State \textbf{At slot $\bm t$, candidate selection:} Each user $i$ becomes a \emph{candidate} if the $\mathbf{q}[kT]$-weighted index value $W^{\mathbf{q}[kT]}_i(\pi_i[t]){>}\omega_{\tau}$, and is not a candidate if \hspace{3pt}$W_i^{\mathbf{q}[kT]}(\pi_i[t])<\omega_{\tau}$.
If $W_i^{\mathbf{q}[kT]}(\pi_i[t])=\omega_{\tau}$, user $i$ becomes a `candidate' with probability $\rho_{\tau}$.
  \State \textbf{At slot $\bm t$, transmission:} If there are no more than $M$ total candidates, then all the candidates are scheduled for data transmission. If there are more than $M$ candidates, then a dummy packet is \emph{broadcasted}.
  \State \textbf{At slot $\bm t$, feedback:} At the end of each slot, if data packets are transmitted, the scheduled users send ARQ feedback to the BS; if the dummy packet is broadcasted, the \emph{candidates} send ARQ feedback to the BS. The belief values are updated correspondingly.
\end{algorithmic}
\end{algorithm}

\noindent \textbf{Remarks}: We next describe the intuition behind designing the above algorithm.

\noindent(1) Note that, for queue stability, instead of using queue length information in every slot, it is sufficient only to consider the sampled queue length information at the periodic slots, i.e., $\mathbf{q}[kT], k=0,1,\cdots$. The queue is stable if and only if the periodically sampled queue length evolution process is stable.

\noindent(2) Within each frame, we wish to maximize the weighted sum-throughput, where each user's throughput is weighted by its queue length sample value at the beginning of the time frame. Hence, in step 2-3, we implement the algorithm $\phi_{str}^{\tau}(\mathbf{q}[kT],N,M,K)$ developed in the previous section. The rationale is because, first, we would like to schedule the users to achieve the higher throughput promised by algorithm $\phi^{\tau}_{str}(\mathbf{q}[kT],N,M,K)$ that exploits the temporal correlated channels. Moreover, for queue stability, we would like to choose users with large queue-lengths.

\noindent(3) Dividing the time slots into different frames brings us advantages in the realm of large frame length $T$. Since we implement the algorithm $\phi_{str}^{\tau}(\mathbf{q}[kT],N,M,K)$ within each finite-horizon frame, if the frame length is small, we lose from exploiting the channel correlation because the optimality of the algorithm requires infinite horizon. As  the frame length scales, the (per-slot) loss of exploiting the channel correlation diminishes.

\vspace{3pt}The next proposition establishes that the throughput region $\bm \Lambda^{N,M}_{rel}$, which is achieved by the optimal policy under a relaxed constraint on the long-term average number of transmissions, can be asymptotically achieved in the stringent constrained scenario by the frame-based algorithm, in the regime of a large number of users. In the proposition, $\mathbf{1}$ is an all $1$ vector, $\tau_0, f(\tau)$ are given in Lemma~\ref{lemma:eps_bound_tau}, and $g(M)$, $\mu(M,K)$ are given in Proposition~\ref{prop:sum_thr_opt}.

\begin{proposition}
\label{prop:region_opt}
{We let $l(M,K)=1-\mu(M,K)$}, if $\tau{\geq} \tau_0$, we have\\
(i) if $K{>}M/2$, for all arrival rate $\mathbf{\bm\lambda}$ with $\mathbf{\bm\lambda}+\big(f(\tau)+2{l}(M,M-g(M))\big) \mathbf{1}\in\bm \Lambda^{N,M}_{rel}$, there exists $T_0$ such that, if $T>T_0$, all queues are stable under the $T$-frame queue-based policy $Frame_{\tau}(T,N,M,M-g(M))$. The function $f(\tau)$ satisfies $\lim_{\tau \rightarrow \infty}f(\tau)=0$.

\noindent(ii) if $\lim_{M\rightarrow\infty}\frac{g(M)}{M}=0$ and $\lim_{M\rightarrow\infty} \frac{g^2(M)}{{M}}=\infty$, then the function ${l}(M,M-g(M))$ satisfies
\begin{align}
\lim_{N\rightarrow \infty}{l}(M,M-g(M))=0.
\end{align}
\end{proposition}

\noindent \textbf{Proof:} We prove the proposition using a Large-Deviation-based Lyapunov technique over time frames. Specifically, we combine the Large Deviation result in Lemma~\ref{lemma:weighted_sum_conv} with uniform convergence of the finite horizon throughput to the infinite horizon throughput performance. We then prove that the average Lyapunov drift of the queue lengths in each time frames is negative, which leads to the stability of the queues. Details of the proof are included in Appendix~.\hfill $\blacksquare$

\vspace{4pt}

\noindent\textbf{Remark:}

\noindent(1) Note that, in Proposition~\ref{prop:region_opt}, the parameter $K$ is kept a distance $g(M)$ from $M$. This mechanism is optimally controls the trade-off between transmitting data packets and broadcasting dummy packets so that we can apply Proposition~\ref{prop:sum_thr_opt} to guarantee the supportable stability region is asymptotic close to the relaxed constrained region $\Lambda^{N,M}_{rel}$, if $g(M)$ scales up at an appropriate rate.

\noindent(2) In the proposed algorithm, a user is selected based on its $\mathbf{q}[kT]$-weighted Whittle's index value in step 3. Since the Whittle's index value measures the importance of scheduling a user under the time-correlated channel, this multiplication captures the importance of scheduling a user under both queue evolution and the time correlation.

\noindent 3) In each frame of algorithm $\text{Frame}_{\tau}(T,N,M,M-g(M))$, implementation of $G^{\tau}(\mathbf{q}[kT],N,M,M-g(M))$ in step $2$ has computational complexity $O((2\tau+1)N \log (2\tau+1)N)$, while implementing step $3$ over the frame has complexity $O(T N)$ (see the remark in Section \ref{sec:thr_appro_alg}). Hence the \emph{per-frame} complexity is $O((2\tau+1)N \log (2\tau+1)N+TN)$. As the frame length $T$ scales up, the \emph{per-slot} complexity decreases toward $O\big(N\big)$.

\section{Conclusion}
In this work, we study downlink scheduling algorithm design over Markovian ON/OFF channels, where the scheduler does not possess accurate instantaneous channel state information. The scheduler instead exploits the Markovian channel memory and channel state feedback from users to make scheduling decisions.
We proposed a low-complexity frame-based algorithm in downlink queuing networks with stringent constraint on the number of simultaneously scheduled users. The proposed algorithm dynamically determines whether to schedule data transmission or broadcast a dummy packet in a slot. By carefully choosing its parameter, the proposed algorithm stably supports arrival rates in a region asymptotically close to that under a relaxed constraint, when the number of users is large. Our on-going work involves comparison of the proposed algorithm with naive/greedy algorithms, as well as designing throughput optimal scheduler under stringent constraint for arbitrary number of users.

\bibliographystyle{IEEEbib}
\bibliography{Bib}

\begin{thebibliography}{1}

\bibitem{MWM}
L. Tassiulas, A. Ephremides,``Dynamic server allocation to parallel
queues with randomly varying connectivity,'' \emph{IEEE Transactions on Information Theory,} vol. 39, pp. 466-478, 1993.

\bibitem{NessLin05}
X. Lin, N. B. Shroff, {``Joint rate control and scheduling
in multihop wireless networks,''} \emph{IEEE CDC}, Atlantis, Bahamas, Dec. 2004.

\bibitem{Eryilmaz05}
A. Eryilmaz, R. Srikant, \emph{``Fair resource allocation in
wireless networks using queue-length based scheduling and congestion
control,''} \emph{IEEE/ACM transactions on networking}, vol. 15, no. 6, pp.
1333-1344, Dec. 2007.

\bibitem{Tse}
D. Tse, P. Viswanath, \emph{``Fundamentals of wireless
communication,''} Cambridge University Press, 2005.

\bibitem{Allerton}
W. Ouyang, S. Murugesan, A. Eryilmaz, N. B. Shroff, ``Scheduling with rate adaptation under incomplete knowledge of channel/estimator statistics,'' in \emph{Allerton Conference,} 2010.

\bibitem{HARQ}
J. Huang, R. A. Berry, and M. L. Honig, ``Wireless scheduling with hybrid ARQ'', IEEE transactins on wireless communicatins, vol. 4, no. 6, 2005.

\bibitem{rohit}
R. Aggarwal, M. Assaad, C. E. Koksal, and P. Schniter,`` Joint scheduling and resource allocation in the ofdma downlink: utility maximization under imperfect channel-state information,'' \emph{IEEE Trans. on Sig. Proc.}, 2011.

\bibitem{Wenzhuo_wiopt13}
W. Ouyang, N. Prasad, S. Rangarajan, ``Exploiting hybrid channel information for downlink multi-user mimo scheduling,'' \emph{IEEE WiOpt 2013}.

\bibitem{WiOpt_version}
W. Ouyang, A. Eryilmaz, N. B. Shroff, ``Low-complexity Optimal Scheduling Over Correlated Fading Channels with ARQ Feedback,'' \emph{IEEE WiOpt 2012}, Paderborn, Germany.

\bibitem{WiOpt_TONversion}
W. Ouyang, A. Eryilmaz, N. B. Shroff, ``Low-complexity Optimal Scheduling Over Correlated Fading Channels with ARQ Feedback,'' under review, \emph{IEEE Transactions on Networking}.

\bibitem{Infocom11}
W. Ouyang, S. Murugesan, A. Eryilmaz, N. Shroff, ``Exploiting channel memory for joint estimation and scheduling in downlink networks,'' \emph{IEEE INFOCOM}, Shanghai, China, Apr. 2011.

\bibitem{Wenzhuo_infocom12}
W. Ouyang, A. Erilmaz, N. B. Shroff, ``Asymptotically optimal downlink scheduling over markovian fading channels,'' \emph{IEEE INFOCOM 2012}.

\bibitem{Javidi}
S.H. Ahmad, M. Liu, T. Javidi, Q. Zhao and B. Krishnamachari,
``Optimality of myopic sensing in multi-Channel opportunistic access,'' \emph{IEEE Trans. on Info. Theory,} 2009.

\bibitem{Zhao_index}
K. Liu, Q. Zhao, ``Indexability of restless bandit problems
and optimality of whittle's index for dynamic multichannel access,''
\emph{IEEE Transactions on Information Theory,} vol. 56, pp. 5547-5567, 2008.

\bibitem{Neely_RR}
C. Li, M. J. Neely, ``Exploiting channel memory for multiuser wireless scheduling without channel measurement: capacity regions and algorithms,'' \emph{Elsevier Performance Evaluation,} 2011.

\bibitem{Neely_RR2}
C. Li, M. J. Neely, ``Network utility maximization over partially observable markovian channels,'' \emph{Elsevier Performance Evaluation}, 2013.

\bibitem{ZhaoTWC}
S.~H.~Ahmad, M.~Liu, T.~Javidi, Q.~Zhao, B.~Krishnamachari, ``Optimality of myopic sensing in
multi-channel opportunistic access,'' \emph{IEEE Transactions on Information Theory,} vol. 55, pp. 4040--4050, 2009.

\bibitem{SM_IT}
S. Murugesan, P. Schniter, N. B. Shroff, ``Multiuser scheduling in a Markov-modeled downlink using randomly delayed ARQ feedback,'' \emph{IEEE Transactions on Information Theory,} vol. 58, no. 2, 2012.

\bibitem{KrishnaModiano}
K. Jagannathan, S. Mannor, I. Menache, E. Modiano, ``A state action frequency approach to throughput maximization over uncertain wireless channels,'' \emph{IEEE INFOCOM}, Shanghai, China, Apr. 2011.

\bibitem{CelicModiano}
G. Celik, E. Modiano, ``Scheduling in networks with time-varying channels and reconfiguration delay,'' \emph{IEEE INFOCOM}, 2012.

\bibitem{Whittle}
P. Whittle, ``Restless bandits: activity allocation in a
changing world,'' \emph{Journal of Applied Probability,} vol. 25, pp. 287-298, 1988.

\bibitem{sugu_aslm}
S. Murugesan, P. Schniter, N. B. Shroff, ``Opportunistic Scheduling using ARQ feedback in Multi-Cell Downlink,'', Asilomar, 2010.


\bibitem{Sondik_thesis}
E. J. Sondik, \emph{``The optimal control of partially observable Markov Decision
Processes,''} PhD thesis, Stanford University, 1971.

\bibitem{Eitan}
Eitan Altman, \emph{``Constrained Markov Decision Processes''}, 1999.

\bibitem{Meyn_CPOMDP}
J. D. Isom, S. Meyn, R. D. Braatz, ``Piecewise linear dynamic programming for constrained POMDPs,'' \emph{National Conference on Artificial Intelligence}, pp. 291-296, 2008.

\bibitem{Neely_tutr}
L. Georgiadis, M. Neely, L. Tassiulas, \emph{``Resource
allocation and cross-Layer control in wireless networks,''} NOW
Publishers Inc., 2006

\bibitem{Weiss_LD}
A. Shwartz, A. Weiss, \emph{``Large deviation for performance analysis,''} Chapman \& Hall, 1994.

\end{thebibliography}

\appendices

\section{Proof of Lemma~\ref{lemma:weighted_sum_conv}}
\label{sec:sum_conv_proof}
Note that
\begin{align}
V_{str}^{\tau}(\mathbf{r},N,M, K)=\sum_{i=1}^N r_i\lim_{T\rightarrow \infty}\frac{1}{T} \mathbb{E}\Big[\sum_{t=0}^{T-1}\hspace{-2pt} \pi_i[t] {\cdot} a_i^{\phi_{str}^{\tau}(\mathbf{r},N,M,K)}[t]\Big], \label{eq:V_str_tau}
\end{align}
where $\pi_i[t]$ evolves according to policy $\phi_{str}^{\tau}(\mathbf{r},N,M, K)$. Next consider the $i$-th summand
\begin{align}
&r_i\lim_{T\rightarrow \infty}\frac{1}{T} \sum_{t=0}^{T-1}\mathbb{E}\Big[\pi_i[t] {\cdot} a_i^{\phi_{str}^{\tau}(\mathbf{r},N,M,K)}[t]\Big] \nonumber\\
=&r_i\lim_{T\rightarrow \infty}\frac{1}{T}\sum_{t=0}^{T-1}\mathbb{E}\Big[ \pi_i[t] {\cdot} \theta_i[t]\cdot \bm 1(\sum_{j\neq i}\theta_j[t]<M) \Big]\nonumber\\
=&r_i\lim_{T\rightarrow \infty}\frac{1}{T}\sum_{t=0}^{T-1}\mathbb{E}\Big[ \pi_i[t]{\cdot}a_i^{\phi_{rel}^{\tau}(\mathbf{r},N,K)}[t]\cdot \bm 1(\sum_{j\neq i}a_i^{\phi_{rel}^{\tau}(\mathbf{r},N,K)}[t]<M) \Big]\nonumber\\
=&r_i\lim_{T\rightarrow \infty}\frac{1}{T}\sum_{t=0}^{T-1}\mathbb{E}\Big[ \pi_i[t]{\cdot}a_i^{\phi_{rel}^{\tau}(\mathbf{r},N,K)}[t]\Big]E\Big[\bm 1(\sum_{j\neq i}a_i^{\phi_{rel}^{\tau}(\mathbf{r},N,K)}[t]<M)\Big].\label{eq:decomp}
\end{align}
where the first equality is from the third step algorithm $\phi_{str}^{\tau}(\mathbf{r},N,M,K)$, where $a_i^{\phi_{str}^{\tau}(\mathbf{r},N,M,K)}[t]=1$ if and only if $\theta_i[t]=1$ and $\sum_{j\neq i}\theta_j[t]<M$. The second equality is because $a_i^{\phi_{rel}^{\tau}(\mathbf{r},N,K)}[t]=\theta_i[t]$, seen from the first remark after the algorithm. The last equality is because each user is scheduled independently under policy ${\phi_{rel}^{\tau}(\mathbf{r},N,K)}$.

Note that, from ergodicity
\begin{align}
\lim_{t\rightarrow \infty} \mathbb{E}\Big[ \pi_i[t]{\cdot}a_i^{\phi_{rel}^{\tau}(\mathbf{r},N,K)}[t]\Big]&= \lim_{T\rightarrow \infty}\frac{1}{T}\sum_{t=0}^{T-1}\mathbb{E}\Big[ \pi_i[t]{\cdot}a_i^{\phi_{rel}^{\tau}(\mathbf{r},N,K)}[t]\Big],\label{eq:erg1}\\
\lim_{t\rightarrow \infty} E\Big[\bm 1(\sum_{j\neq i}a_i^{\phi_{rel}^{\tau}(\mathbf{r},N,K)}[t]<M) \Big]&=\lim_{T\rightarrow \infty}\frac{1}{T}\sum_{t=0}^{T-1}E\Big[\bm 1(\sum_{j\neq i}a_i^{\phi_{rel}^{\tau}(\mathbf{r},N,K)}[t]<M)\Big].\label{eq:erg2}
\end{align}

Therefore, from (\ref{eq:decomp})-(\ref{eq:erg2}) we have
\begin{align}
&r_i\lim_{T\rightarrow \infty}\frac{1}{T} \sum_{t=0}^{T-1}\mathbb{E}\Big[\pi_i[t] {\cdot} a_i^{\phi_{str}^{\tau}(\mathbf{r},N,M,K)}[t]\Big] \nonumber\\
=&r_i\lim_{T\rightarrow \infty}\frac{1}{T}\sum_{t=0}^{T-1}\mathbb{E}\Big[ \pi_i[t]{\cdot}a_i^{\phi_{rel}^{\tau}(\mathbf{r},N,K)}[t]\Big]\cdot\lim_{T\rightarrow \infty}\frac{1}{T}E\Big[\bm 1(\sum_{j\neq i}a_i^{\phi_{rel}^{\tau}(\mathbf{r},N,K)}[t]<M) \Big]\nonumber\\
=&r_i\lim_{T\rightarrow \infty}\frac{1}{T}\sum_{t=0}^{T-1}\mathbb{E}\Big[ \pi_i[t]{\cdot}a_i^{\phi_{rel}^{\tau}(\mathbf{r},N,K)}[t]\Big]\cdot\lim_{t\rightarrow \infty}Pr(\sum_{j\neq i}a_i^{\phi_{rel}^{\tau}(\mathbf{r},N,K)}[t]<M).\label{eq:str_rel_connect}
\end{align}

Therefore we have
\begin{align}
V_{str}^{\tau}(\mathbf{r},N,M, K)=&\sum_{i=1}^N r_i\lim_{T\rightarrow \infty}\frac{1}{T} \mathbb{E}\Big[\sum_{t=0}^{T-1}\hspace{-2pt} \pi_i[t] {\cdot} a_i^{\phi_{str}^{\tau}(\mathbf{r},N,M,K)}[t]\Big]\nonumber\\
=& \sum_{i=1}^N r_i\lim_{T\rightarrow \infty}\frac{1}{T}\sum_{t=0}^{T-1}\mathbb{E}\Big[ \pi_i[t]{\cdot}a_i^{\phi_{rel}^{\tau}(\mathbf{r},N,K)}[t]\Big]\cdot\lim_{t\rightarrow \infty}Pr(\sum_{j\neq i}a_i^{\phi_{rel}^{\tau}(\mathbf{r},N,K)}[t]<M)\nonumber\\
\leq& \sum_{i=1}^N r_i\lim_{T\rightarrow \infty}\frac{1}{T}\sum_{t=0}^{T-1}\mathbb{E}\Big[ \pi_i[t]{\cdot}a_i^{\phi_{rel}^{\tau}(\mathbf{r},N,K)}[t]\Big]\nonumber\\
=&V_{rel}^{\tau}(\mathbf{r},N,K)\nonumber
\end{align}
which proves the second inequality in (\ref{eq:str_rel_bound}).

We let $a_i^{\phi_{rel}^{\tau}(\mathbf{r},N,K)}[\infty]$ be a random variable, which has the same distribution with the stationary distribution of $a_i^{\phi_{rel}^{\tau}(\mathbf{r},N,K)}[t]$. Since $a_i^{\phi_{rel}^{\tau}(\mathbf{r},N,K)}[\infty]\geq 0$, we have
\begin{align}
Pr\big(\sum_{i\neq j} a_i^{\phi_{rel}^{\tau}(\mathbf{r},N,K)}[\infty]<M\big)\geq Pr\big(\sum_{i=1}^N a_i^{\phi_{rel}^{\tau}(\mathbf{r},N,K)}[\infty]<M\big).\label{eq:prob_bound}
\end{align}

We next bound the right hand side of the above inequality.

\begin{lemma}\label{lemma:LD_bound}
If $K>M/2$, \begin{align}
Pr\big(\sum_{i=1}^N a_i^{\phi_{rel}^{\tau}(\mathbf{r},N,K)}[\infty]<M\big)\geq 1-\exp\Big(-\frac{(M-K)^2}{3K}\Big). \nonumber
\end{align}
\end{lemma}

\textbf{Proof:} See Appendix~\ref{sec:LD_sqrt}.\hfill$\blacksquare$

Therefore
\begin{align}
V_{str}^{\tau}(\mathbf{r},N,M,K)=&\sum_{i=1}^N r_i\lim_{T\rightarrow \infty}\frac{1}{T} \sum_{t=0}^{T-1}\mathbb{E}\Big[\pi_i[t] {\cdot} a_i^{\phi_{str}^{\tau}(\mathbf{r},N,M,K)}[t]\Big]\nonumber\\
\geq &\Big(1-\exp\big(-\frac{(M-K)^2}{3K}\big)\Big) \sum_{i=1}^N r_i\lim_{T\rightarrow \infty}\frac{1}{T}\sum_{t=0}^{T-1}\mathbb{E}\Big[ \pi_i[t]{\cdot}a_i^{\phi_{rel}^{\tau}(\mathbf{r},N,K)}[t]\Big]\nonumber\\
=& \Big(1-\exp\big(-\frac{(M-K)^2}{3K}\big)\Big)V_{rel}^{\tau}(\mathbf{r},N,K).
\end{align}

From Lemma~\ref{lemma:LD_bound}, and (\ref{eq:V_str_tau})(\ref{eq:str_rel_connect})(\ref{eq:prob_bound}) we have
\begin{align}
1\geq\frac{V_{str}^{\tau}(\mathbf{r},N,M,K)}{V_{rel}^{\tau}(\mathbf{r},N,M)}&=\frac{V_{str}^{\tau}(\mathbf{r},N,M,K)}{V_{rel}^{\tau}(\mathbf{r},N,K)}\cdot \frac{V_{rel}^{\tau}(\mathbf{r},N,K)}{V_{rel}^{\tau}(\mathbf{r},N,M)}\nonumber\\
&\geq \Big(1-\exp\big(-\frac{(M-K)^2}{3K}\big)\Big)\frac{V_{rel}^{\tau}(\mathbf{r},N,K)}{V_{rel}^{\tau}(\mathbf{r},N,M)}\label{eq:1stMulti}
\end{align}

If the total number of user $i$ with non-negative weights, i.e., with $r_i>0$, is no more than $K$, then all the users are scheduled to transmit at each slot in both policy $\phi_{rel}^{\tau}(\mathbf{r},N,M)$ and $\phi_{rel}^{\tau}(\mathbf{r},N,K)$, we hence have
\begin{align}
V_{rel}^{\tau}(\mathbf{r},N,M)= V_{rel}^{\tau}(\mathbf{r},N,K).\nonumber
\end{align}

Now consider the scenario where the total number of user $i$ with $r_i>0$ is more than $K$. We define the set $\Theta=\{i:Pr\big(a_i^{\phi_{rel}^{\tau}(\mathbf{r},N,M)}[\infty]=1\big)>0\}$, i.e., the set $\Theta$ consists the index of all users that contributes to the steady state throughput under policy $\phi_{rel}^{\tau}(\mathbf{r},N,M)$. Hence $|\Theta|
\geq K$. We order the indices in $\Theta$ so that $r_{\sigma(1)}\leq r_{\sigma(2)}\leq \cdots \leq r_{\sigma(|\Theta|)}$.

We let $\chi=\min\{M,|\Theta|\}$ and $x$ be such that $\sum_{i=1}^{x} Pr\big(a_i^{\phi_{rel}^{\tau}(\mathbf{r},N,M)}[\infty]=1\big)=\chi-K$. Now consider another heuristic policy $\tilde{\phi}_{rel}^{\tau}(\mathbf{r},N,M)$ which is exactly the same as policy $\phi_{rel}^{\tau}(\mathbf{r},N,M)$ except that user $1, \cdots, x$ are no longer scheduled. The time-average amount of users scheduled under $\tilde{\phi}_{rel}^{\tau}(\mathbf{r},N,M)$ is hence $K$. Therefore, he long-term average throughput $\widetilde{V}_{rel}^{\tau}(\mathbf{r},N,M)$ of policy $\tilde{\phi}_{rel}^{\tau}(\mathbf{r},N,M)$ satisfies
\begin{align}
\widetilde{V}_{rel}^{\tau}(\mathbf{r},N,M)&\geq {V}_{rel}^{\tau}(\mathbf{r},N,M)-(M-K) r_{\sigma(x)},\label{eq:tilde2rel}\\
{V}_{rel}^{\tau}(\mathbf{r},N,K)&\geq\widetilde{V}_{rel}^{\tau}(\mathbf{r},N,M).\label{eq:rel2tilde}
\end{align}

From (\ref{eq:tilde2rel})-(\ref{eq:rel2tilde}) we have
\begin{align}
{V}_{rel}^{\tau}(\mathbf{r},N,K)\geq {V}_{rel}^{\tau}(\mathbf{r},N,M)-(M-K)r_{\sigma(x)}.\label{eq:km_diff}
\end{align}

Note that, under the policy $\tilde{\phi}_{rel}^{\tau}(\mathbf{r},N,M)$, the total number of users, in steady state, that contributes to throughput equals $|\Theta|-x$, therefore $|\Theta|-x\geq K$. Now consider another policy $\hat{\phi}_{rel}^{\tau}(\mathbf{r},N,K)$ that only schedules the $K$ users with the highest weights, i.e., users $\sigma(|\Theta|), \sigma(|\Theta|-1),\cdots,\sigma(|\Theta|-K+1)$. Therefore the corresponding long-term average throughput satisfies
\begin{align}
\widehat{V}_{rel}^{\tau}(\mathbf{r},N,K)\geq& \sum_{i=|\Theta|-K+1}^{|\Theta|} r_{\sigma(i)}b^{\sigma(i)}_s \nonumber\\
\geq& \delta \sum_{i=|\Theta|-\chi+1}^{|\Theta|} r_{\sigma(i)}\nonumber\\
\geq& \delta (\chi-1) r_{\sigma(|\Theta|-\chi+1)}\nonumber\\
\geq& \delta (K-1) r_{\sigma(x)}. \nonumber
\end{align}

Therefore
\begin{align}
V_{rel}^{\tau}(\mathbf{r},N,M)\geq \widehat{V}_{rel}^{\tau}(\mathbf{r},N,M)\geq \delta (K-1)r_{\sigma(x)}.\label{eq:relMbound}
\end{align}

From (\ref{eq:km_diff}) and (\ref{eq:relMbound}) we have
\begin{align}
\frac{V_{rel}^{\tau}(\mathbf{r},N,M,K)}{V_{rel}^{\tau}(\mathbf{r},N,M)}=&\frac{V_{rel}^{\tau}(\mathbf{r},N,M)-(M-K)r_{\sigma(x)}}{V_{rel}^{\tau}(\mathbf{r},N,M)}\nonumber\\
\geq&\Big(1-\frac{(M-K)r_{\sigma(x)}}{V_{rel}^{\tau}(\mathbf{r},N,M)}\Big)\nonumber\\
\geq& \Big(1-\frac{(M-K)r_{\sigma(x)}}{\delta (K-1)r_{\sigma(x)}}\Big)\nonumber\\
\geq& \Big(1-\frac{M-K}{\delta (K-1)}\Big).\label{eq:2ndMulti}
\end{align}

Substituting (\ref{eq:2ndMulti}) in (\ref{eq:1stMulti}), we have
\begin{align}
1\geq& \frac{V_{str}^{\tau}(\mathbf{r},N,M,K)}{V_{rel}^{\tau}(\mathbf{r},N,M)}=\frac{V_{str}^{\tau}(\mathbf{r},N,M,K)}{V_{rel}^{\tau}(\mathbf{r},N,M,K)}\cdot \frac{V_{rel}^{\tau}(\mathbf{r},N,M,K)}{V_{rel}^{\tau}(\mathbf{r},N,M)}\nonumber\\
\geq& \Big[1-\exp(-\frac{(M-K)^2}{3K})\Big]\Big[1-\frac{M-K}{\delta (K-1)}\Big]^+.\nonumber
\end{align}

We let $\mu(M,K)=\Big[1-\exp(-\frac{(M-K)^2}{3K})\Big]\Big[1-\frac{M-K}{\delta (K-1)}\Big]^+$.
The Lemma is thus proven.

\section{Proof of Proposition~\ref{prop:region_opt}}

Define Lyapunov function $L(\mathbf{q})=\frac{1}{2}\sum_{i=1}^N q_i^2$. We consider \emph{the $T$-frame average Lyapunov drift $\Delta L(\mathbf{q}[kT])$ over the $k$-th frame}, expressed as,
\begin{align}
&\Delta L(\mathbf{q}[kT])/T\nonumber\\
=&\frac{1}{T} \mathbb{E} \Big[L(\mathbf{q}[(k+1)T])- L(\mathbf{q}[kT]) \big | \ \mathbf{q}[kT], \bm \pi[kT] \Big]\nonumber\\
\leq & BT+ \sum_{i=1}^N q_i[kT] \cdot \lambda_i - \sum_{i=1}^N q_i[kT]\cdot\frac{1}{T} \nonumber\\
&\hspace{-8pt}\cdot\mathbb{E}\Big[ \sum_{t=0}^{T-1} \pi_i[kT{+}t] {\cdot} a_i^{\phi_{str}^{\tau}\big(\mathbf{q}[kT],N,M,M-g(M)\big)}[kT{+}t] \Big | \bm \pi[kT] \Big],\label{eq:drift}
\end{align}
where $B$ is a constant whose value is determined by the second moment of the arrival process \cite{Neely_tutr}. Because $\bm \lambda+\big(f(\tau)+2l(M,M-g(M))\big) \mathbf{1} \in \bm\Gamma$, for any non-negative vector $\mathbf{q}$, we have
\begin{align}
\sum_{i=1}^N q_i \cdot (\lambda_i +\big(f(\tau)+2l(M,M-g(M))\big)) \leq V_{rel}^*(\mathbf{q},N,M),\nonumber
\end{align}
where $V_{rel}^*(\mathbf{q}[kT],N,M)$ is defined in (\ref{eq:thr_nontrun}).
The Lyapunov drift (\ref{eq:drift}) now becomes,
\begin{align}
\Delta L(\mathbf{q}[kT])/T&\leq BT{-}\big(f(\tau)+2l(M,M-g(M))\big)\sum_{i=1}^N q_i[kT]{+}\nonumber \\
&V_{rel}^*(\mathbf{q}[kT],N,M){-}V^{\tau,T}_{str}(\mathbf{q}[kT],N,M,M{-}g(M))\nonumber\\
&\hspace{-0.9in}= BT{-}\big(f(\tau)+2l(M,M-g(M))\big)\sum_{i=1}^N q_i[kT]{+} V_{rel}^*(\mathbf{q}[kT],N,M){-}V_{rel}^{\tau}(\mathbf{q}[kT],N,M)\nonumber \\
&\hspace{-0.7in}+V_{rel}^{\tau}(\mathbf{q}[kT],N,M){-}V_{str}^{\tau}(\mathbf{q}[kT],N,M,M{-}g(M))\nonumber\\
&\hspace{-0.7in}+V_{str}^{\tau}(\mathbf{q}[kT],N,M{-}g(M)){-}V^{\tau,T}_{str}(\mathbf{q}[kT],N,M,M{-}g(M)).\label{eq:drift_decomp}
\end{align}
where $V_{rel}^{\tau}(\mathbf{q}[kT],N,M)$  and $V_{str}^{\tau}(\mathbf{q}[kT],N,M{-}g(M))$ are defined in (\ref{eq:thr_trun}) and (\ref{eq:thr_tru_str}), and $V^{\tau,T}_{str}(\mathbf{q}[kT],N,M,M{-}g(M))$ is the $T$-horizon expected transmission rate achieved under the policy $\phi_{str}^{\tau}\big(\mathbf{q}[kT], N, M,M-g(M)\big)$, i.e.,
\begin{align}
\hspace{-6pt}&V^{\tau,T}_{str}(\mathbf{q}[kT],N,M,M{-}g(M))\nonumber\\
{=}&\sum_{i=1}^N q_i[kT]\frac{1}{T} \mathbb{E}\Big[ \sum_{t=0}^{T-1}\hspace{-2pt} \pi_i[kT{+}t] {\cdot} a_i^{\phi_{str}^{\tau}\big(\mathbf{q}[kT], N,M,M-g(M)\big)}[kT{+}t] \Big | \bm \pi[kT] \Big]. \nonumber
\end{align}

Note that, in~(\ref{eq:drift_decomp}), the difference $V_{rel}^*(\mathbf{q}[kT],N,M){-}V_{rel}^{\tau}(\mathbf{q}[kT],N,M)$ is bounded in Lemma~\ref{lemma:eps_bound_tau} as follows,
\begin{align}\label{eq:tau_conv}
V_{rel}^*(\mathbf{q}[kT],N,M){-}V_{rel}^{\tau}(\mathbf{q}[kT],N,M)\leq f(\tau)\cdot \sum_{i=1}^N q_i[kT].
\end{align}

The difference $V_{rel}^{\tau}(\mathbf{q}[kT],N,M){-}V_{str}^{\tau}\big(\mathbf{q}[kT],N,M,M{-}g(M)\big)$ is bounded in Lemma~\ref{lemma:weighted_sum_conv} as
\begin{align}
V_{rel}^{\tau}(\mathbf{q}[kT],N,M){-}V_{str}^{\tau}\big(\mathbf{q}[kT],N,M,M{-}g(M)\big)\leq& [1-\mu\big(M,M{-}g(M)\big)] V_{str}^{\tau}\big(\mathbf{q}[kT],N,M,M{-}g(M)\big)\nonumber\\
=& l(M,M{-}g(M)) V_{str}^{\tau}\big(\mathbf{q}[kT],N,M,M{-}g(M)\big)\label{eq:K_conv}
\end{align}

The following bound is from \cite{WiOpt_version}\cite{WiOpt_TONversion}, which states that, as the length of the time horizon tends to infinity, the expected achieved rate in finite horizon asymptotically converges to infinite horizon achievable rate.

\vspace{8pt}\begin{lemma}
\label{lemma:exp_decay}
For any M and $\kappa>0$, we have, uniformly over $\mathbf{q}$, $M$, and the initial state $\bm \pi[kT]$, there exist positive constants $c_1$ and $c_2$ such that
\begin{align}
\Big|V_{str}^{\tau}\big(\mathbf{q},N,M,M{-}g(M)\big){-}V^{\tau,T}_{str}\big(\mathbf{q},N,M,M{-}g(M)\big)\Big| < \big(\kappa{+}c_1 \exp({-}c_2 T) \big) \sum_{i=1}^N q_i. \nonumber
\end{align}
\end{lemma}

\vspace{5pt}

From Lemma~\ref{lemma:eps_bound_tau} and (\ref{eq:tau_conv})(\ref{eq:K_conv}), the Lyapunov drift (\ref{eq:drift_decomp}) can be further bounded as follows,
\begin{align}
&\Delta L(\mathbf{q}[kT])/ T \nonumber\\
\leq & BT{+}\nonumber\\
&\hspace{0.1in}\Big[{-}\big(f(\tau)-2l(M,M-g(M))+f(\tau)+l(M,M-g(M))+\big(\kappa+c_1 \exp(-c_2 T) \big)\Big]\cdot \sum_{i=1}^N q_i[kT] \nonumber \\
=& BT{+}\Big[{-}l(M,M-g(M)){+}\big(\kappa{+}c_1 \exp({-}c_2 T) \big)\Big] \sum_{i=1}^N q_i[kT]. \label{eq:drift_bound}
\end{align}
For fixed $\tau$, by choosing $\kappa$ sufficiently small and $T$ sufficiently large, say $T>T_0$, the Lyapunov drift is negative whenever the sum of the queue lengths gets sufficiently large. Therefore, the queues are stable according to the Foster-Lyapunov criterion. Part (ii) of the proposition follows directly from Proposition~\ref{prop:sum_thr_opt}.

\section{Proof of Lemma~\ref{lemma:LD_bound}}
\label{sec:LD_sqrt}

The proof is in line with Large Deviation principles \cite{Weiss_LD}. Note that traditional Large Deviation techniques, in our context, holds for linear growth of $K$. Here, instead, we identify the growth rate of $K$ that leads to our desired result.

For notational convenience, we use to $a_i$ represent $a_i^{\phi_{rel}^{\tau}}(\mathbf{r},N,K)[\infty]$. Note that, from Lemma~\ref{lemma:eps_bound_tau}(ii), we have
\begin{align}
\sum_{i=1}^N \mathbb{E}[a_i]\leq K. \label{eq:mean}
\end{align}

From Markov's inequality, we have for arbitrary $t\geq 0$,
\begin{align}
Pr\Big(\sum_{i=1}^N a_i\geq M\Big)\leq& \frac{\mathbb{E}\Big[\exp\big(t\sum_{i=1}^N a_i\big)\Big]}{\exp\Big(t\cdot M\Big)}\nonumber\\
=& \frac{\prod_{i=1}^N \mathbb{E}\Big[\exp\big(t\cdot a_i\big)\Big]}{\exp\Big(t\cdot M\Big)}\nonumber\\
=&\frac{\prod_{i=1}^N \Big[\Big(1-Pr\big(a_i=1\big)\Big)+Pr\big(a_i=1\big)e^t\Big]}{\exp\Big(t\cdot M\Big)}\nonumber\\
=&\frac{\prod_{i=1}^N \Big[1+Pr\big(a_i=1\big)\big(e^t-1\big)\Big]}{\exp\Big(t\cdot M\Big)}\nonumber\\
\leq&\frac{\prod_{i=1}^N \exp{\Big[Pr\big(a_i=1\big)\big(e^t-1\big)\Big]}}{\exp\Big(t\cdot M\Big)}\nonumber\\
=&\frac{\exp\Big[(e^t-1)\mathbb{E}\big[\sum_{i=1}^N a_i \big]\Big]}{\exp\Big(t\cdot M\Big)}\nonumber\\
\leq &\frac{\exp\Big[(e^t-1)K\Big]}{\exp\Big(t\cdot M\Big)}\nonumber\\
=& \exp\big(\eta(t)\big),\label{eq:chernoff}
\end{align}
where the first inequality is from Markov's inequality, and the second inequality is because $1+x\leq e^x$ for $x\geq 0$, and the last inequality is from (\ref{eq:mean}). The function $\eta(t)$ is defined as follows,
\begin{align}
\eta(t)=\exp\Big[\exp\big[(e^t-1)K\big]-t\cdot M\Big].\label{eq:eta_t}
\end{align}

We let $t^*$ to be the minimal point of $\eta(t)$, i.e., $\eta'(t^*)=0$, we then have
\begin{align}
t^*=\log\Big(\frac{M}{K}\Big).
\end{align}

Therefore
\begin{align}
Pr\Big(\sum_{i=1}^N a_i\geq M\Big)\leq\exp\big(\eta(t^*)\big).\nonumber
\end{align}

Substituting the expression of $t^*$ to $\eta'(t^*)$ in (\ref{eq:eta_t}), we have
\begin{align}
\eta(t^*)=&\exp\Big[\big(\frac{M}{K}-1\big)K-M\cdot \log\big(\frac{M}{K}\big)\Big]\nonumber\\
=&\exp\Big[M-K-M\cdot \log\big(\frac{M}{K}\big)\Big]\nonumber\\
=& \exp\Big[\big(M-K\big)-K\big(1+\frac{M-K}{K}\big)\cdot \log\big(1+\frac{M-K}{K}\big)\Big]\label{eq:delta_bound}
\end{align}

Note that, for $0\leq\delta< 1$, we have $\log(1+\delta)=\sum_{n=1}^{\infty}(-1)^{n+1}\frac{x^n}{n}$ and hence
\begin{align}
(1+\delta)\log(1+\delta)=&\delta+\sum_{n=2}^{\infty}(-1)^n\delta^i\Big(\frac{1}{n-1}-\frac{1}{n}\Big)\nonumber\\
\geq& \delta+\frac{1}{2}\delta^2-\frac{1}{6}\delta^3\nonumber\\
\geq& \delta+\frac{1}{3}\delta^2.\label{eq:delta_bound2}
\end{align}

Since $K> M/2$, we have $(M-K)/K<1$. From~(\ref{eq:delta_bound}) and (\ref{eq:delta_bound2}),
\begin{align}
\eta(t^*)\leq&\exp\Big[M-K-K\big(\frac{M-K}{K}+\frac{(M-K)^2}{3K^2}\big)\Big]\nonumber\\
=&\exp\Big[-\frac{(M-K)^2}{3K}\Big].
\end{align}

From (\ref{eq:chernoff}) we have
\begin{align}
Pr\Big(\sum_{i=1}^N a_i< M\Big)&\geq Pr\Big(\sum_{i=1}^N a_i< M\Big)\nonumber\\
&\geq 1-\exp\big(\eta(t^*)\big)\nonumber\\
&\geq \exp\Big[-\frac{(M-K)^2}{3K}\Big],\nonumber
\end{align}
which proves the lemma.

\end{document}